\documentclass[useAMS,usenatbib,usegraphicx,referee]{mn2e}
\title[Sky-distribution of Gamma-Ray Bursts]{Testing the Randomness
in the Sky-Distribution of Gamma-Ray Bursts}
\author[R. Vavrek et al.]{R. Vavrek,$^{1}$ \thanks{E-mail: Roland.Vavrek@sciops.esa.int (RV)}
L.~G.~Bal\'azs,$^{2}$ A.~M\'esz\'aros,$^{3}$ I.~Horv\'ath$^{4}$ and Z.~Bagoly$^{5}$\\
$^{1}$ESA/ESAC P.O. Box 50727
Villafranca del Castillo, 28080 Madrid, Spain\\
$^{2}$Konkoly Observatory, P.~O.~Box 67, H-1525 Budapest, Hungary\\
$^{3}$Astronomical Institute of the Charles University, V Hole\v{s}ovi\v{c}k\'ach 2, 180 00 Prague 8, Czech Republic\\
$^{4}$Department of Physics, Bolyai Military University,  P.~O.~Box 15, H-1581 Budapest, Hungary\\
$^{5}$Laboratory for Information Technology, E\"{o}tv\"{o}s
University, P\'azm\'any P\'eter s\'et\'any 1/A, H-1518 Budapest,
Hungary}

\begin{document}

\date{Accepted XXXX ... XX. Received XXXX ... XX; in original form XXXX ... XX}

\pagerange{\pageref{firstpage}--\pageref{lastpage}} \pubyear{2007}

\maketitle

\label{firstpage}

\begin{abstract}
We studied the complete randomness of the angular distribution of
gamma-ray bursts (GRBs) detected by BATSE. Since GRBs seem to be a
mixture of objects of different physical nature we divided the
BATSE sample into 5 subsamples (short1, short2, intermediate,
long1, long2) based on their durations and peak fluxes and studied
the angular distributions separately. We used three methods,
Voronoi tesselation, minimal spanning tree and multifractal
spectra to search for non-randomness in the subsamples. To
investigate the eventual non-randomness in the subsamples we
defined 13 test-variables (9 from the Voronoi tesselation, 3 from
the minimal spanning tree and one from the multifractal spectrum).
Assuming that the point patterns obtained from the BATSE
subsamples   are fully random we made Monte Carlo simulations
taking into account the BATSE's sky-exposure function. The MC
simulations enabled us to test the null hypothesis i.e. that the
angular distributions are fully random. We tested the randomness
by binomial test and introducing  squared Euclidean distances in
the parameter space of the test-variables. We concluded that the
short1, short2 groups deviate significantly (99.90\%, 99.98\%)
from the fully randomness in the distribution of the squared
Euclidean distances but it is not the case at the long samples. At
the intermediate group the squared Euclidean distances  also give
significant deviation (98.51\%).
\end{abstract}

\section{Introduction}
Recently, there is no doubt about the cosmological origin of the
gamma-ray bursts (hereafter GRBs) \citep{zame04,fox05,me06}. Then,
assuming a large scale isotropy for the Universe, one expects the
same property for the GRBs as well. Another property, which is
also expected to occur that GRBs should appear fully randomly,
i.e.  if a burst is observed it does not give any information
about the place of the next one. If both properties are fulfilled,
then the distribution is called completely random (for the
astronomical context of spatial point processes see
\citet{pato95}). There are several tests for checking the complete
randomness of point patterns, however, these procedures do not
always give information for both properties simultaneously.

There are increasing evidence that all the GRBs do not represent a
physically homogeneous group
\citep{kou93,ho98,muk98,hak00,ho02,bal03,hak03,ho06}.
 Hence, it is worth investigating that the physically different subgroups are
also different in their angular distributions. In the last years
the authors provided \citep{ba98,ba99,me00a,me00b}  several
different tests probing the intrinsic isotropy in the angular
sky-distribution of GRBs collected in BATSE Catalog \citep{mee00}.
Shortly summarizing the results of these studies one may conclude:
A. The long subgroup ($T_{90} > 10\; s$) seems to be distributed
isotropically; B. The intermediate subgroup ($2\;s \lid T_{90}
\lid 10\;s $) is distributed anisotropically on the $\simeq
(96-97)$\% significance level; C. For the short subgroup ($2\; s >
T_{90}$) the assumption of isotropy is rejected only on the $92$\%
significance level; D. The long and the short subclasses,
respectively, are distributed differently on the $99.3$\%
significance level.  (About the definition of subclasses see
\citet{ho98}; $T_{90}$ is the duration of a GRB, during which time
the $90$\% of the radiated energy is received \citep{mee00}.)

Independently and by different tests, \citet{li01}  confirmed the
results A., B. and C. with one essential difference: for the
intermediate subclass a much higher - namely $99.89$\% -
significance level of anisotropy is claimed. Again, the short
subgroup is found to be "suspicious", but only the $\simeq
(85-95)$\% significance level is reached. The long subclass seems
to be distributed isotropically (but see \citet{mest03}).
\citet{magl03} found significant angular correlation on the
$2^{\circ} - 5^{\circ}$ scale for GRBs with $T_{90}< 2s$
durations. \citet{tan05}  reported a correlation between the
locations of previously observed short bursts and the positions of
galaxies in the local Universe, indicating that between 10 and 25
per cent of short GRBs originate at low redshifts ($z < 0.025$).

It is a reasonable requirement to continue these tests using more
sophisticated procedures in order to see whether the angular
distribution of GRBs is completely random or has some sort of
regularity. This is the subject of this article. New tests will be
presented here. Mainly the clarification of the short subgroup's
behaviour is expected from these tests. In this paper, similarly
to the previous studies, the {\it intrinsic\/} randomness is
tested; this means that the non-uniform sky-exposure function of
BATSE instrument is eliminated.

The paper is organized as follows. In Section \ref{mat} the three
new tests are described. This Section does not contain new
results, but - because the methods are not widely familiar - this
minimal survey may be useful. Section \ref{tests} contains the
statistical tests on the data. Section  \ref{disc} summarizes the
results of the statistical tests,  and Section \ref{conc} presents
the main conclusions of the paper.

\section{Mathematical summary}
\label{mat}
\subsection{Voronoi tesselation (VT)}
\label{vt}
 The Voronoi diagram - also known as Dirichlet
tesselation or Thiessen polygons - is a fundamental structure in
computational geometry and arises naturally in many different
applications \citep{vor,stoy}. Generally, this diagram provides a
partition of a point pattern ("point field", also "point process")
according to its spatial structure, which can be used for
analyzing  the underlying point process.

Assume that there are $N$ points ($N \gg 1$) scattered on a sphere
surface with an unit radius. One says that a point field is given
on the sphere. The Voronoi cell \citep{stoy} of a point is the
region of the sphere surface consisting of points which are closer
to this given point than to any other ones of the sphere. This
cell forms a polygon on this sphere. Every such cell has its area
($A$) given in steradian, perimeter ($P$) given by the length of
boundary (one great circle of the boundary curve is called also as
"chord"), number of vertices ($N_v$) given by an integer positive
number, and by the inner angles ($\alpha_i;\; i = 1,..., N_v$).
This method is completely non-parametric, and therefore may be
sensitive for various point pattern structures in the different
subclasses of GRBs.

Note that the behaviour of this tesselation method on the sphere
surface is quite different from that on the infinite plane. This
follows from the fact that the polygon's area will not be
independent from each other, because the total  surface of the
sphere  is fixed in $4\pi$ steradian. Hence, the spherical Voronoi
tesselation is not effected by border effects, and the Voronoi
diagram becomes a closed set of convex polygons.

The points on sphere may be distributed  completely randomly or
non-randomly; the non-random distribution may have different
characters (clustering, filaments, etc.; for the survey of these
non-random behaviours see, e.g., \citet{dig}).

Random and some regular patterns have  distributions of one
characteristic maxima (unimodal) but with different variances. The
multimodality means different characteristic maxima indicating
hierarchical (cluster) structure, the number of modes is
determined by the number of scales in the sample. The VT method is
able both to detect the non-randomness  and to describe its form
(for more details see \citet{stoy} and for the astronomical
context \citet{col90,col91,ick91,ike91, sub92,
wey94,zan95,doros97,yaha99,ram01}).

\begin{figure}
\centering
 \includegraphics[width=94mm]{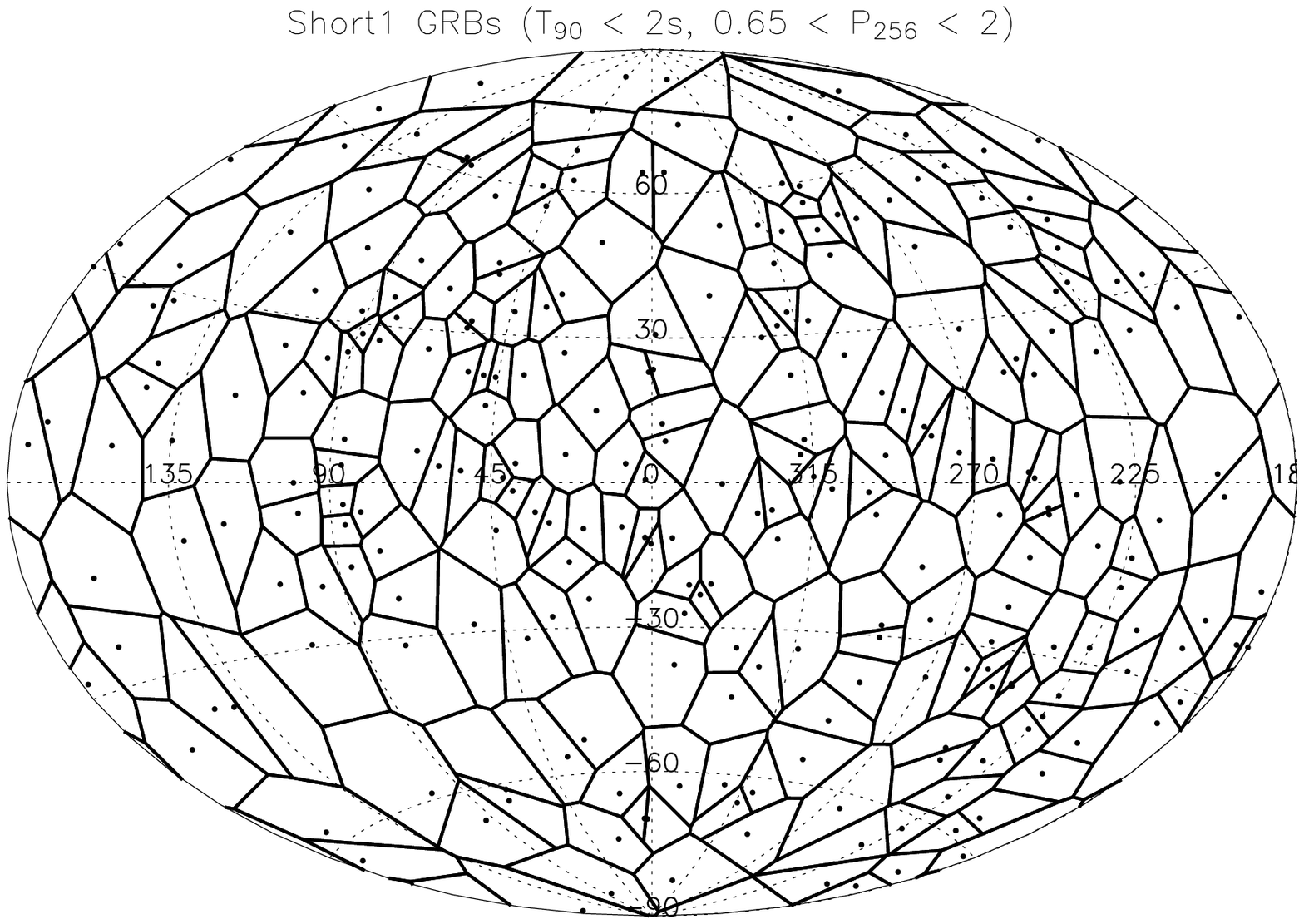}
 \caption{Voronoi tesselation of the short GRBs (Short1 sample)
 in the $0.65 < P_{256} < 2.00$
 peak-flux range in Galactic coordinates.
  The peak-flux is given in dimesion photon/(cm$^2$s).}
  \label{vors}
\end{figure}

\subsection{Minimal spanning tree (MST)}
\label{mist}
 Contrary to VT, this method considers
 the distances (edges) among the
points (vertices). Clearly, there are $N(N-1)/2$ distances among
$N$ points. A spanning tree is a system of lines connecting all
the points without any loops. The minimal spanning tree (MST) is a
system of connecting lines, where the sum of the lengths is
minimal among all the possible connections between the points
 \citep{kru56,prim57}.  In this paper the spherical
version of MSF is used following the original Prim's paper.

The $N-1$ separate connecting lines (edges) together define the
minimal spanning tree. The statistics of the lengths and the
$\alpha_{MST}$ angles between the edges at the vertices  can be
used for testing the randomness of the point pattern. The MST is
widely used in cosmology for studying the statistical properties
of galaxy samples
\citep{bar85,bha96,krze96,bhav96,adam99,doros01}.

\begin{figure}
\centering
  \includegraphics[width=94mm]{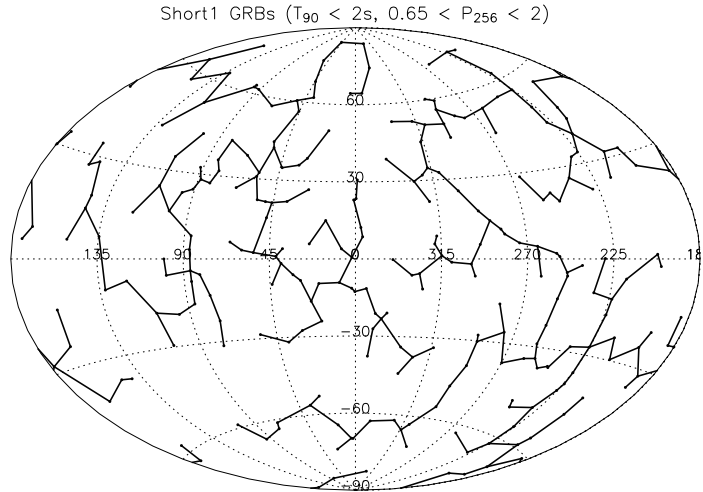}
  \caption{The MST for the sample in Fig.\ref{vors}.}
  \label{mst}
\end{figure}

\subsection{Multifractal spectrum}
\label{mfs}
 Let denote $P(\varepsilon)$ the probability for
finding a point in an area of $\varepsilon$ radius.  If
$P(\varepsilon)$ scales as $\varepsilon^{\alpha}$ (i.e.
$P(\varepsilon)\propto \varepsilon^{\alpha}$), then $\alpha$ is
called the local fractal dimension (e.g. $\alpha=2$ for a
completely random process on the plane). In the case of a
monofractal $\alpha$ is independent on the position. A
multifractal (MFR) on a point process can be defined as
unification of the subsets of different (fractal) dimensions
\citep{pala}. One usually denotes with $f(\alpha)$ the fractal
dimension of the subset of points at which the local fractal
dimension is in the interval  of $\alpha,\alpha+d\alpha$. The
contribution of these subsets to the whole pattern is not
necessarily equally weighted, practically it depends on the
relative abundances of subsets. The $f(\alpha)$ functional
relationship between the fractal dimension of subsets and the
corresponding local fractal dimension is called the MFR or
Hausdorff spectrum.

In the vicinity of $i$-th point ($i=1,2,...,N$) one can measure
from the neighbourhood structure a local dimension $\alpha_i$
("R\'enyi dimension"). This measure approximates the dimension of
the embedding subset, giving a possibility to construct the MFR
spectrum which characterizes the whole pattern (for more details
see \citet{pala}). If the maximum of this convex spectrum is equal
to the Euclidean dimension of the space, then in classical sense
the pattern \emph{is not a fractal}, but the spectrum remains
sensitive to the non-randomness of the point set.

There is a  wide variety of astronomical phenomena, where the
concept of fractal and/or multifractal can be successfully applied
\citep{gir00,irw00,kav00,
pan00,sel00,bot01,cel01,cha01,tat01,vav01,asch02,cas02,elm02,gai02,
pan02,sem02,tik02,dat03}.

\begin{figure}
\centering
  \includegraphics[width=94mm]{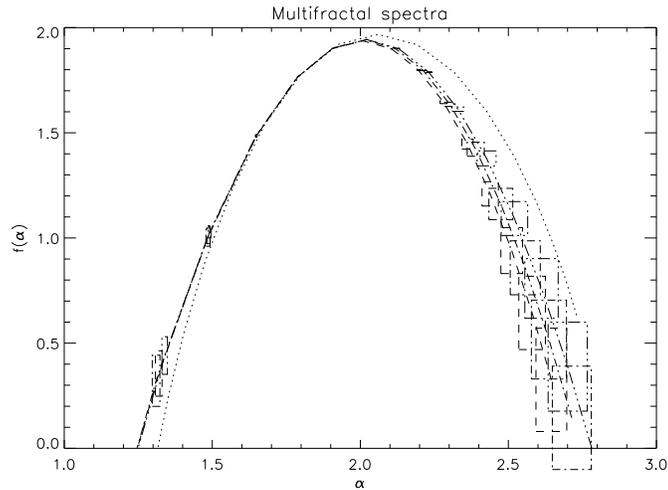}
  \caption{MFR spectra of simulated (dot-dashed), Long1 (dashed),
   Short1 (dotted) and Short2 (three-dot-dashed) samples. Boxes represent
   the error of spectrum points derived from Monte Carlo simulations.
   Note the shift of the maximum of the spectrum of the
   Short1 sample towards higher values in comparison to $\alpha=2$,
   corresponding to the completely random 2D Euclidean case.}
  \label{rmfr}
\end{figure}

\section{Statistical tests on the data}
\label{tests}
 The three procedures outlined in Section \ref{mat}
 enable  us to derive several stochastic quantities well suited for
testing the non-randomness of the underlying point patterns.

\subsection{Input data and the definition of samples}
\label{input}
Up to the present the most comprehensive all-sky
survey of GRBs was done by the BATSE experiment on board of the
CGRO satellite in the period of 1991-2000. In this period the
experiment collected 2704 well justified burst events and the data
are available in the Current BATSE Catalog \citep{mee00}.

Since there are increasing evidence (\citet{ho06} and references
therein) that the GRB population is actually a mixture of
astrophysically different phenomena, we  divided the GRBs into
three groups: $short$ ($T_{90} < 2s$), $intermediate$ ($2s\leq
T_{90} \leq 10s$) and $long$ ($T_{90} > 10s$). To avoid the
problems with the changing detection threshold we omitted the GRBs
having a  peak flux $P_{256} \le 0.65 \ photons \ cm^{-2} \
s^{-1}$. This truncation was proposed by \citet{pen97}. The bursts
may emerge at very different distances in the line of sight and it
may happen that the stochastic structure of the angular
distribution depends on it. Therefore, we also made tests on the
bursts with $P_{256} < 2\ photons \ cm^{-2} \
s^{-1}$ in the short and long population, separately. 
Table \ref{test} defines the 5 samples to be studied here.

\begin{table}
\centering
  \caption{Tested samples of BATSE GRBs.}
  \label{test}
  \begin{tabular}{@{}lllc}
  \hline
  Sample & Duration  & Peak flux  & Number  \\
         &    [$s$]    &  [$photons$ $cm^{-2}$$s^{-1}$] & of GRBs   \\
  \hline
  Short1  &  $T_{90}<2$ s & $0.65 < P_{256} < 2$ & 261 \\
  Short2  &  $T_{90}<2$ s & $0.65 < P_{256}$     & 406 \\
  Intermediate   & $2\;s \lid T_{90} \lid 10\;s $ & $0.65 < P_{256}$ & 253 \\
  Long1   &  $T_{90}>2$ s & $0.65 < P_{256} < 2$ & 676 \\
  Long2   &  $T_{90}>10$ s & $0.65 < P_{256}$    & 966 \\
  \hline
  \end{tabular}
\end{table}

\subsection{Definition of the test-variables}
\label{deftest}
 The randomness of the point field on the sphere
 can be tested with respect to different criteria. Since
different non-random behaviours are sensitive for different types
of criteria of non-randomness, it is not necessary that  all
possible tests using different measures reject the assumption of
randomness. In the following we defined several test-variables
which are sensitive to different stochastic properties of the
underlying point pattern, as proposed by \citet{wadu98}.

\subsubsection{Voronoi tesselation}
Any of the four quantities characterizing the Voronoi cell, i.e.
the area, the perimeter, the number of vertices, and the inner
angles can be used as test-variables or even some of their
combinations, too. We defined the following test-variables:

\begin{itemize}
\item[-]Cell area  $ A $;

\item[-]Cell vertex (edge)   $ N_v $;

\item[-]Cell chords    $ C $;

\item[-]Inner angle  $ \alpha_i$;

\item[-]Round factor (RF) average $RF_{av}=\overline{4\pi A / P
}$;

\item[-]Round factor (RF) homogeneity  $1-\frac{\sigma
(RF_{av})}{RF_{av}}$;

\item[-]Shape factor   $ A/P^2$;

\item[-]Modal factor   $\sigma (\alpha_i) / N_v$;

 \item[-]The so-called "AD factor"
defined as $AD = 1 - (1 - \sigma (A) / \langle A \rangle )^{-1}$,
where $\sigma (A)$ is the dispersion and $\langle A \rangle$ is
the average of $A$.
\end{itemize}

\subsubsection{Minimal spanning tree}
To characterize the stochastic properties of a point patters we
use three quantities obtained from a MST:

\begin{itemize}

\item[-]Variance of the MST edge-length
$\sigma(L_{MST})$;

\item[-]Mean MST edge-length   $L_{MST}$;

\item[-]Mean  angle between edges   \ $\alpha_{MST}$.
\end{itemize}

\subsubsection{Multifractal spectrum}
Here the only used variable is the  $f(\alpha)$ multifractal spectrum, which is
 is a sensitive tool for testing the non-randomness of a point pattern.

 An important problem is to study the sensitivity (discriminant
power) of the different parameters  to the different kind of
regularity inherent in the point pattern. In the case of a fully
regular mesh, e.g., $A$ is constant and so $AD=0$, $\sigma(
\alpha_i)=0$ and both are increasing towards a fully random
distribution. In case of a patchy pattern the distribution of the
area of the Voronoi cells and the edge distribution of MST become
bimodal reflecting the average area and edge length within and
between the clusters, in comparison to the fully random case. In a
filamentary distribution the shape of the areas becomes strongly
distorted reflecting in an increase of the modal factor in
comparison to the case of patches.

\citet{wadu97} investigated the power of the Voronoi tesselation
and minimal spanning tree in discriminating between distributions
having big and small clusters, full randomness and hard cores
(random distributions but the mutual distances of the points are
constrained by the size of a hard core), respectively. They
concluded that  Voronoi roundness factor did not separate small
clusters and hardcore distributions, and roundness factor
homogeneity did not distinguish between small clusters and random
distributions, {\bf nor} between random and hardcore
distributions. MST has a very good discriminant power even in the
case of hardcore distributions with close minimal interpoint
distances.

Since the sensitivity of the variables are different on  changing
the regularity properties of the underlying point patterns one may
measure significant differences in one parameter but not in the
other even in the case when these are correlated otherwise. It is
not a trivial issue. In most cases one needs extended numerical
simulations to study the statistical significance of the different
parameters.

\subsection{Estimation of the significance}
\label{empro} Let us denote with $\xi$ one of the thirteen
test-variables defined in Section \ref{deftest}. The probability
that $\xi<x$ occurs  is given by $P(\xi<x)= F(x)$, where $F(x)$
is the probability distribution function. We approximated  $F(x)$
numerically by the $F_n(x)$ empirical probability distribution
function which can be calculated by $F_n(x)=k/n$ where $n$ is the
number of simulations and $k$ is the number of cases for which the
simulated $\xi<x$ holds.

Similarly, the probability that $\xi$ is within the interval
$[x_1;x_2]$ can be obtained by $F_n(x_1)-F_n(x_2)=(x_2-x_1)/n$.
Then the $\beta$ probability that $\xi$ is outside this region is
given by $\beta=1-(x_2-x_1)/n$. In the following we suppose that
the $[x_1;x_2]$ interval is symmetric to the $\overline{x}$ sample
mean. To obtain the empirical distributions of the test-variables
we made 200 simulations for each of the five samples. The number
of the simulated points were identical with those of the samples
defined in Section \ref{input}.

We generated the fully random catalogs by Monte Carlo (MC)
simulations of fully random GRB celestial positions and taking
into account the BATSE sky-exposure function
 (\cite{fish}, \cite{mee00}).

Assuming that the point patterns obtained from the five samples,
defined in Table 1, are fully random we calculated
the 
 probabilities for all the 13 test-variables selected
in Section \ref{deftest}.  Based on the simulated distributions we
computed the level of significance for all the 13 test-variables
and in all samples defined.

\section{Discussion of the statistical properties}
\label{disc}
\subsection{Significance of independent multiple tests}
\label{bnom} In Section \ref{empro}  we calculated numerical
significance for the tests assuming they were performed
individually. The calculated significance levels are given in
Table~\ref{testtab}. In the reality, however, these figures would
mean significance at a certain level if one performed only that
single test. Assuming that all the single tests were independent
the $P_n(m)$ probability that among $n$ trials at least $m$ will
resulted significance only by chance at a certain level is given
by the following equality:

\begin{equation}
\label{btest}
 P_n(m) = \sum \limits^n_{k=m} P^n_k \,,
\end{equation}

\noindent where $P^n_k$ is the binomial distribution giving the
probability of $k$ successes among $n$ trials:

\begin{equation}
\label{binom}
 P^n_k = \frac{n!}{k!(n-k)!}p^k(1-p)^{n-k}.
\end{equation}

\noindent In the equation given above $p$ means the probability
that a single test has given significant result only by chance. It
is easy to see that this equation resulted in $P_n(1) = 1 -
(1-p)^n \approx np$ which gives a significance of $1-np$
approximately instead of $1-p$. It means e.g. that a single test
resulted in $1-p=0.95$ significance is reduced to
$1-0.95^{13}=0.49$ if one performed $n=13$ independent tests but
only one resulted in $1-p=0.95$ significance.

Inspecting Table \ref{testtab} listing the  calculated numerical
significance of single tests one can infer that $short1$ sample
has 4 tests with significance of $1-p=0.95$ ore more.  Taking into
account the calculations at the end of the previous paragraph,
however, we have to emphasize that the calculated numerical
significance, based on the individual probability distribution of
the test-variables separately, does not have its original meaning.
Significance refers to the certainty rejecting the null hypothesis
on the basis of the bunch of tests as a whole. Applying Equation
(\ref{btest}) with $m=4$ and $n=13$ one gets a significance of
99.69~\%. Applying the same sequence of arguments to the $short2$
sample one may get a figure of only 86.46~\% ($m=2$ and $n=13$).
In the case of the $intermediate$, $long1$ and $long2$ samples one
can not get figures above the 95~\% significance level.

One may have a serious concern, however, with the results obtained
above. Namely, the basic requirement of the independence of the
single tests is not fulfilled in our case. In the contrary, there
are strong correlations between the test-variables in Table
\ref{testtab}. In the next subsection we try to outline an
approach which takes into account the correlations between the
test-variables and overcomes this difficulty.

\begin{table}
\centering \caption{Calculated significance levels for the 13
test-variables and the five samples. A {\bf calculated numerical}
significance greater than 95\% is put in bold
face.}\label{testtab}
\begin{tabular}{llcccccc}
\hline
 Name &  var & short1  & short2 & interm. &  long1  &  long2\\
 \hline
Cell area  &  $ A $ &  36.82  &   29.85  &   94.53  &  79.60 &   82.59\\
Cell vertex (edge)  & $ N_v $  &  36.82  &   87.06  &  \ 2.99  &  26.87 &  \ 7.96\\
Cell chords  &  $ C $  &  47.26  &   52.24  &   18.91  &  84.58 &   54.23\\
Inner angle  &  $ \alpha_i $  &  {\bf 96.52}  &   21.39  &   87.56  &  37.81 &   63.18\\
RF average &  $\overline{4\pi A / P }$  &  65.17  &  {\bf 99.98}  &   33.83  &  10.95 &   86.07 \\
RF homogeneity & $1-\frac{\sigma (RF_{av})}{RF_{av}}$   &  19.90  &   24.38  &   58.71  &  55.72 &   32.84 \\
Shape factor  &  $ A/P^2 $  &  91.04  &   94.03  &   90.05  &  55.22 &   63.68\\
Modal factor &  $\sigma (\alpha_i) / N_v$  &  {\bf 97.51}  &  \ 1.99  &  \ 7.46  &  56.22 &  \ 8.96\\
AD factor &  $1- \bigl(1- \frac{\sigma (A)} { \langle A \rangle}\bigr)^{-1}$   &  32.84  &   25.37  &   11.44  &  {\bf 95.52} &   52.74 \\
MST variance  & $\sigma(L_{MST})$     & 52.74  &   38.31  &   22.39  &  13.93 &   59.70\\
MST average  &  $L_{MST} $     & {\bf 97.51}  &  \ 7.46  &   89.05  &  56.72 &  \ 8.96\\
MST angle   & \ $\alpha_{MST}$     & 85.07  &   14.43  &   36.82  &  73.63 &   60.70\\
MFR spectra &  $f(\alpha)$      & {\bf 95.52}  &   {\bf 96.02}  &   {\bf 98.01}  &  73.63 &   36.32\\
\hline Binomial test & (Eq. (\ref{btest}) with $p=0.05$) &
{\bf99.69} & 86.46 &
51.33 & 51.33 & -  \\
 \hline
Squared Euclidean & distance &  {\bf 99.90} & {\bf 99.98}  & {\bf 98.51} & 93.03  &   36.81       \\
\hline
\end{tabular}
\end{table}

\subsection{Evaluation of the joint significance levels}
We assigned  to every MC simulated sample 13 values of the test
variables  and, consequently, a point in the 13D parameter space.
Completing 200 simulations in all of the subsamples we get in this
way  a 13D sample representing the joint probability distribution
of the 13 test-variables. Using a suitable chosen measure of
distance of the points from the sample mean we can get a
stochastic variable characterizing the deviation of the simulated
points from the mean only by chance. An obvious choice would be
the squared Euclidean distance.

In case of a Gaussian distribution with unit variances and without
correlations this would resulted in a  $\chi^2$ distribution of 13
degree of freedom. The test-variables in our case are correlated
and have different scales. Before computing squared Euclidean
distances we transformed the test-variables into non-correlated
ones with unit variances. Due to the strong correlation between
some of the test-variables we may assume that the observed
quantities can be represented with non-correlated variables of
less in number. Factor analysis (FA) is a suitable way to
represent the correlated observed variables with non-correlated
variables of less in number.

Since our test-variables are stochastically dependent following
\citet{wadu98} we attempted to represent them by fewer
non-correlated hidden variables supposing that

\begin{equation}
\label{fac}
 X_i = \sum \limits^k_{j=1} a_{ij} f_j + s_i
\phantom{@@@@@} i=1, \ldots,p \ ; \phantom{@@} k<p \ .
\end{equation}

\noindent In the above  equation $X_i, f_j, s_i$ mean the
test-variables ($p=13$ in our case), the hidden variables and a
noise-term, respectively. Equation (\ref{fac}) represents the
basic model of FA.  After making some reasonable assumptions
\citep{KS73}, $k$ can be constrained by the following inequality:

\begin{equation}
\label{constr}
 k \leq (2p+1- \sqrt{8p+1})/2
\end{equation}

\noindent which gives $k \leq 8.377$ in our case.

Factor analysis is a common ingredient of professional statistical
software packages (BMDP, SAS, S-plus, SPSS\footnote{BMDP, SAS,
S-plus, SPSS are registered trademarks}, etc).  The default
solution for identifying the factor model is to perform principal
component analysis (PCA).  We kept as many factors as were
meaningful with respect to Equation (\ref{constr}). Taking into
account the constraint imposed by Equation~(\ref{constr}) we
retained 8 factors. In this way we projected the joint
distribution of the test-variables in the 13D parameter space into
a 8D one defined by the non-correlated $f_i$ hidden variables.

The $f_j$ hidden variables in Equation (\ref{fac}) are
non-correlated and have zero means and unit standard deviations.
Using these variables we defined the following squared Euclidean
distance from the sample mean:

\begin{equation}
\label{eudist}
 d^2 = f_1^2 + f_2^2 + \ldots + f_k^2 \phantom{@@@@}
(k=8 \ \textrm{ in our case}) \ .
\end{equation}

 \noindent If the $f_j$ variables had strictly Gaussian
 distributions
 Equation (\ref{eudist}) would define a $\chi^2$ variable of $k$ degrees of
 freedom.

\begin{figure}
\centering
  \includegraphics[width=124mm]{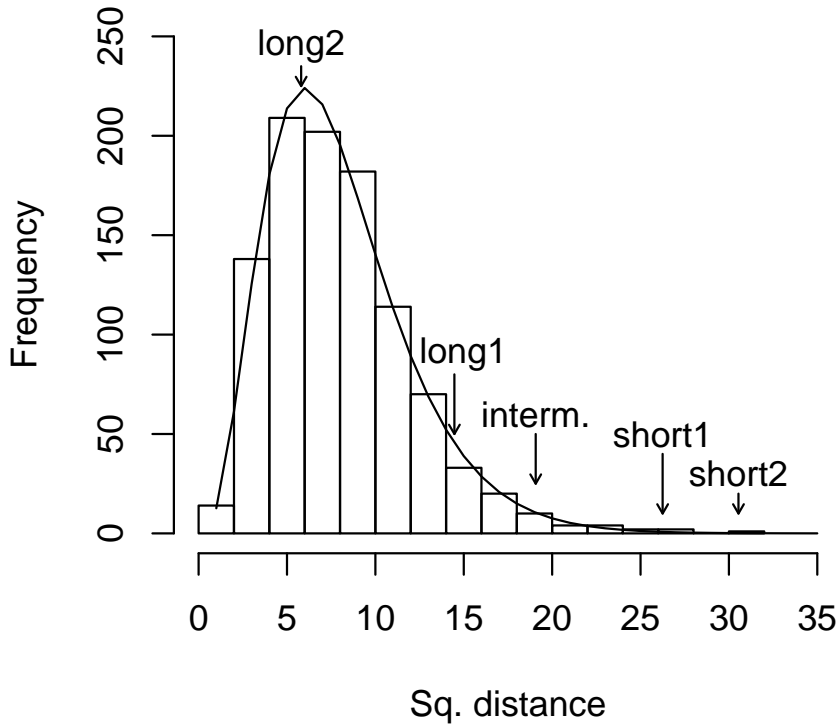}
  \caption{Distribution of the squared Euclidean distances of the simulated samples
  from the stochastic mean of the $f_i$ hidden variables (factors) in the 8D parameter
  space. There are altogether 1000 simulated points. Full line marks
  a $\chi^2$ distribution of 8 degree of freedom, normalized to the sample size.
  The distances of the BATSE samples
  are also indicated. The departures of samples $"short1"$ and
  $"short2"$ exceed all those of the simulated points. The
  probabilities, that these deviations are non-random, equal
  99.9\% and 99.98\%, respectively.
}
  \label{hist}
\end{figure}

\subsection{Statistical results and their interpretations}
In addition to the significance obtained by the binomial test in
Subsection \ref{bnom} using the distribution of the squared
Euclidean distances, defined by Equation (\ref{eudist}),   one can
get further information  whether a BATSE sample represented by a
point in the parameter space of the test-variables deviates only
by chance or it significantly differs from the fully random
distribution.

In all categories ($short1$, $short2$, $intermediate$, $long1$,
$long2$) we made 200, altogether 1000, simulations. We calculated
the $d^2$ squared distances for all simulations and compared them
with those of the BATSE samples in Table \ref{test}. Figure
\ref{hist} shows a histogram of the simulated squared distances
along with those of the BATSE samples. Full line represent  a
$\chi^2$ distribution of $k=8$ degree of freedom. Figure
\ref{hist} clearly shows that the departures of samples $short1$
and $short2$ exceed all those of the simulated points. The
probabilities, that these deviations are non-random, equal 99.9\%
and 99.98\%, respectively.

The full randomness of the angular distribution of the long GRBs,
in contrast to the regularity of the short and  in some extent to
the intermediate ones, points towards the differences in the
angular distribution of their progenitors. The recent discovery of
the afterglow in some short GRBs  indicates that these events are
associated with the old stellar population \citep{fox05} accounted
probably for the mergers of compact binaries, in contrast to the
long bursts resulting from  the collapse of very massive stellar
objects in young star forming regions. The differences in
progenitors reflects also the differences between the energy
released by the short and long GRBs.

Unfortunately, little can be said on the physical nature of the
intermediate class.  The statistical studies (\citet{ho06}
 and the references therein) suggest the existence of this
subgroup - at least from the purely statistical point of view.
Also the non-random sky distribution is occurring here. But its
physical origin is fully open yet \citep{ho06}.

\section{Summary and conclusions}
\label{conc} We made additional studies on the degree of the
randomness in the angular distribution of samples selected from
the BATSE Catalog. According to the $T_{90}$ durations and
$P_{256}$ peak fluxes of the GRBs in the Catalog we defined five
groups: $short1$ ($T_{90}<2$ s \& $0.65 < P_{256} < 2$), $short2$
($T_{90}<2$ s \& $0.65 < P_{256}$ ), $intermediate$ ($2\;s \lid
T_{90} \lid 10\;s $ \& $0.65 < P_{256}$), $long1$ ($T_{90}>2$ s \&
$0.65 < P_{256} < 2$) and $long2$ ($T_{90}>10$ s \& $0.65 <
P_{256}$).

To characterize the statistical properties of the point patterns,
given by the samples, we defined 13 test-variables based on the
Voronoi tesselation (VT), Minimal spanning tree (MST) and
Multifractal spectra. For all five GRB samples defined we  made
200 numerical simulations assuming fully random angular
distribution and taking into account the BATSE exposure function.
The numerical simulations enabled us to define empirical
probabilities  for testing the null hypothesis, i.e. the
assumption that the angular distributions of the BATSE samples are
fully random.

 Since we performed 13 single tests simultaneously on each
subsamples the significance obtained by calculating it separately
for each test can not be treated as a true indication for
deviating from the fully random case. At first we supposed that
the test-variables were independent and making use the binomial
distribution computed the probability of obtaining significant
deviation  in at least one of the variables only by chance. In
fact, some of the test-variables are strongly correlated. To
concentrate the information on the non-randomness experienced by
the test-variables, we assumed that they can be represented as a
linear combination of non-correlated hidden factors of less in
number. Actually, we estimated $k=8$ as the number of hidden
factors. Making use the hidden factors we computed the
distribution of the squared Euclidean distances from the mean of
the simulated variables. {\it Comparing  the distribution of the
squared Euclidean distances of the simulated with the BATSE
samples we concluded that the short1, short2 groups deviate
significantly (99.90\%, 99.98\%)  from the fully randomness, but
it is not the case at the long samples. At the intermediate group
 squared Euclidean distances also give significant deviation
(98.51\%).}

\section*{Acknowledgments}

This study was supported by OTKA grant No. T048870, by a Bolyai
Scholarship (I.H.), by a Research Program MSM0021620860 of the
Ministry of Education of Czech Republic, and by a GAUK grant No.
46307 (A.M.). We are indebted to an an anonymous referee for his
valuable comments and suggestions.

\bsp

\label{lastpage}

\end{document}